\documentclass{article}
\usepackage{natbib,epsfig,emulateapj}

\newcommand{\rhob}{\rho_{\rm b}}

\begin{document}
\submitted{To appear in ApJ Letters}

\lefthead{Bildsten \& Ushomirsky}
\righthead{Viscous Boundary Layer Damping of R-Modes in Neutron Stars} 

\title{Viscous Boundary Layer Damping of R-Modes in Neutron Stars} 

\author{Lars Bildsten} 
\affil{Institute for Theoretical Physics and Department of Physics,\\
         Kohn Hall, University of California, Santa Barbara, CA 93106
	\\ bildsten@itp.ucsb.edu}

\and

\author{Greg Ushomirsky}
\affil{Department of Physics and Department of Astronomy,\\ 
601 Campbell Hall, University of California, Berkeley, CA 94720 \\
gregus@fire.berkeley.edu}

\begin{abstract}

Recent work has raised the exciting possibility that r-modes (Rossby
waves) in rotating neutron star cores might be strong gravitational
wave sources. We estimate the effect of a solid crust on their viscous
damping rate and show that the dissipation rate in the viscous
boundary layer between the oscillating fluid and the nearly static
crust is $>10^5$ times higher than that from the shear throughout the
interior. This increases the minimum frequency for the onset of the
gravitational r-mode instability to at least 500 Hz when the core
temperature is less than $10^{10} \ {\rm K}$. It eliminates the
conflict of the r-mode instability with the accretion-driven spin-up
scenario for millisecond radio pulsars and makes it unlikely that the
r-mode instability is active in accreting neutron stars.  For newborn
neutron stars, the formation of a solid crust shortly after birth
affects their gravitational wave spin-down and hence detectability by
ground-based interferometric gravitational wave detectors.

\end{abstract}


\section{Introduction} 
\label{sec:introduction}

 The recently discovered r-mode instability
\citep{andersson98:_new_class,friedman98:_axial_instab} might play a
significant role in setting the spin frequencies ($\nu_s$) of rapidly
rotating neutron stars
\citep{lindblom98:_gravit_radiat_instab,owen98:_gravit,andersson99:_gravit_radiat,bildsten98:gravity-wave,andersson99:accreting_rmode}.
The unstable regime for the r-modes in the neutron star (NS) core
depends on the competition between the gravitational radiation
excitation and viscous
dissipation. \citet{lindblom98:_gravit_radiat_instab} and
\citet{andersson99:_gravit_radiat} computed the dissipation due to
shear and bulk viscosities in normal fluids (not superfluids) and
found that gravitational excitation exceeds viscous damping in
all but the most slowly rotating stars. \citet{lindblom99:superfluid}
showed that superfluid mutual friction is also not competitive with
gravitational radiation unless the superfluid entrainment parameter
assumes a very special value. Hence, on theoretical grounds, r-modes
are expected to be unstable over much of the parameter space occupied
by newborn NSs, NSs in low-mass X-ray binaries (LMXBs), and
millisecond radio pulsars (MSPs) during spin-up.

However, observations pose challenges to this theoretical picture. The
existence of two $1.6$~ms radio pulsars means that rapidly rotating
NSs are formed in spite of the r-mode instability
\citep{andersson99:accreting_rmode}.  While it is not clear that their
{\it current} core temperatures place these MSPs inside the r-mode
instability region, current theory says that they were certainly
unstable during spin-up. For accreting NSs in LMXBs,
\citet{bildsten98:gravity-wave} and
\citet{andersson99:accreting_rmode} conjectured that the gravitational
radiation from an r-mode with a small constant amplitude could balance
the accretion torque.  This would possibly explain the preponderance
of $\approx 300$~Hz spins among LMXBs \citep{klis99:_millis}, which
would otherwise have been spun up to $>1000 \ {\rm Hz}$ during their
$\gtrsim10^9$~yr lifetime \citep{vanparadijs95:lmxb_distrib}.
However, for normal fluid cores, \citet{levin99} showed that the
temperature dependence of the shear viscosity makes this spin
equilibrium thermally unstable, leading to a limit cycle behavior of
rapid spin-downs after prolonged periods of spin-up.  For superfluid
cores, \citet{bu99:rmodes} showed that such constant-amplitude spin-up
is inconsistent with the quiescent luminosities of several known
LMXBs.

 All but the hottest NSs have solid crusts that occupy the outer
$\approx $~km of the star.  In this paper, we calculate a previously
overlooked dissipation: the viscous boundary layer between the
oscillating fluid core and static crust. The shear viscosity of NS
matter is relatively small and does not affect the r-mode structure in
the stellar interior. However, the transverse fluid motions of the
r-modes are very large at the crust-core boundary and must, of course,
go to zero relative velocity at the crust. The dissipation in the
resulting viscous boundary layer (hereafter referred to as the VBL)
substantially shortens (by typically $10^5$) the r-mode damping times.
Previous work
\citep{lindblom98:_gravit_radiat_instab,Lindblom99:2ndorder,andersson99:_gravit_radiat}
included damping from the shear viscosity acting only on the gradient
in the transverse velocity of the mode in the interior of the star (on
the length scale of the stellar radius, $R$).  The shear in the VBL
acts on a much shorter length scale, and is hence stronger.

 This new source of dissipation raises the minimum frequency for the
r-mode instability in NSs with a crust to $\gtrsim500$~Hz for
$T\approx10^{10}$~K, and even higher frequencies for lower
temperatures. It thus alleviates the conflict between the 
accretion-driven spin-up scenario for MSPs and the r-mode instability and
significantly alters the spin-down scenario proposed by
\citet{owen98:_gravit} for newborn NSs.

\section{The Viscous Boundary Layer}
\label{sec:boundary-layer}

  The transverse fluid motions of an r-mode cause a time-dependent
``rubbing'' of the core fluid against the otherwise co-rotating crust
spinning at $\Omega=2\pi \nu_s$. In previous works, the boundary
condition applied at this location allowed the fluid to have
large-amplitude transverse motion. Neglecting viscosity is an
excellent assumption far away from the crust-core interface.  However,
there can be no relative motion at the boundary for a viscous fluid,
leading to a VBL where the transverse velocity drops from a large
value to zero (the ``no-slip'' condition). The VBL is mediated by the
kinematic viscosity, $\nu$, in the fluid just beneath the base of the
crust. The density there has most recently been estimated as
$\rhob\approx 1.5\times10^{14}{\rm~g~cm}^{-3}$ \citep{pethick95}.  We
use the values for $\nu$ as found by \citet{flowers79} and fit by
\citet{cutler87},
\begin{equation}\label{eq:viscosity}
\nu=1.8\times 10^4 {\rm~cm}^2 {\rm~s}^{-1}\frac{f}{T_8^{2}}.
\end{equation}
In this fit, three different cases are handled with the parameter $f$.
When both neutrons and protons are normal, the predominant scatterers
are neutrons, and $f=(\rho/\rhob)^{5/4}$. If neutrons are superfluid
and protons normal, the viscosity is mediated by electron-proton
scattering, and $f=1/15$. Finally, when both protons and neutrons are
superfluid, electron-electron scattering yields $f=5 (\rho/\rhob)$.

  We begin by assuming that the crust is infinitely rigid, and hence
the waves do not penetrate into the crust.  In an oscillating flow,
the thickness of the VBL, $\delta$, is found by setting the
oscillation frequency of the r-mode in the rotating frame,
$\omega=2m\Omega/l(l+1)=2\Omega/3$ ($l=m=2$), to the inverse of the
time it takes vorticity to diffuse across its width, i.e. 
$\delta^2/\nu\sim \omega^{-1}$.  In the plane-parallel case,
\citet{landau59:_fluid_mech} show that the correction to the velocity
in the VBL of an incompressible oscillatory flow falls off
exponentially with a length
\begin{equation}\label{eq:bl-thickness}
\delta=\left(\frac{2\nu}{\omega}\right)^{1/2}\approx 
3{\rm~cm~}\frac{f^{1/2}}{T_8}
\left(\frac{1{\rm~kHz}}{\nu_s}\right)^{1/2},
\end{equation}
much smaller than $R$.  This estimate neglects the Coriolis
force. When the VBL is determined by the balance between the Coriolis
force, $\sim2\Omega v$, and viscosity, $\sim\nu v/\delta^2$ (the Ekman
problem), then the thickness is $\delta\sim(\nu/\Omega)^{1/2}$,
roughly the same as eq.~(\ref{eq:bl-thickness}).  Hence, while the
Coriolis force will change the angular dependence of the velocity in
the VBL, it is unlikely to change its thickness.

The viscous dissipation can be handled quite simply for this thin VBL. 
Since all of the kinetic energy there is damped in one
cycle, the $Q$ of the oscillation is roughly just the ratio of the
total volume to that in the VBL, or $Q\sim R/3\delta\sim 10^5 T_8
(\nu_s/1{\rm~kHz})^{1/2} /f^{1/2}$. The damping time is then a few
hundred seconds, thus competitive with the growth time from
gravitational wave emission, $\tau_{\rm gw}= -146{\rm~s~}
M_{1.4}^3 R_6^{-9} (\nu_s / 1{\rm~kHz})^{-6}$, where $M_{1.4}=M/1.4
M_\odot$ and $R_6=R/10 {\rm \ km}$, of the $l=m=2$ r-mode
\citep{owen98:_gravit,lindblom98:_gravit_radiat_instab}.

To calculate the damping more accurately, we use the dissipation per
unit area from \citet{landau59:_fluid_mech} and integrate over the
surface area at the crust-core boundary
\begin{equation}\label{eq:dEdt}
\frac{dE}{dt}=-\int\frac{\rho v^2}{2}
\left(\frac{\omega\nu}{2}\right)^{1/2} R^2 \sin\theta d\theta d\phi.
\end{equation}
The fluid velocity is given by $\vec v=(\alpha\Omega R)\vec
Y_{lm}^Be^{i\omega t}$, and $E=1.64\times10^{-2} M (\alpha\Omega
R)^2/2$ is the mode energy \citep{owen98:_gravit}.  Integrating and
using the mode frequency in the rotating frame, we find the damping
time due to rubbing
\begin{equation}\label{eq:tau-rub}
\tau_{\rm{rub}}=-\frac{2E}{dE/dt} \nonumber \\
	\approx100{\rm~s~}
	\frac{M_{1.4} T_8}{R_6^2 f^{1/2}}
	\left(\frac{\rhob}{\rho}\right)
	\left(\frac{1{\rm~kHz}}{\nu_s}\right)^{1/2},
\end{equation}
where $\rho$ is
the density at the crust-core boundary. 

So far, we have assumed that the large-scale relative motion between
the oscillating core and the crust is simply set by the toroidal
motion of the r-mode. This is fine in the inviscid limit
($\nu=0$). The transverse displacement $\xi_\perp$ at the crust-core
boundary is then discontinuous, and the r-modes do not couple to the
crust's torsional modes, which have $\vec{\xi}=\xi_\perp
\vec{Y}_{lm}^B$. For $\nu=0$, r-modes can only couple to the crust via
their $\mathcal{O}(\Omega^2)$ radial motion.  However, the VBL exerts a
time-dependent shear stress, $\sigma_{r\perp}= (\omega\nu)^{1/2}\rho
v$ \citep{landau59:_fluid_mech}, that shakes the crust and can
potentially couple the crust to the core.  If this coupling can drive
the amplitude of the crustal motion $\xi_\perp$ to be comparable to
the transverse motions in the r-modes ($\xi_\perp=3\alpha R/2$ for
$l=m=2$), then our picture of the core rubbing against a static crust
would need to be reconsidered.  The energy in an r-mode oscillation is
$\sim10^{51}\alpha^2 (\nu_s/1{\rm~kHz})^2{\rm~ergs}$, while the
typical energy in torsional oscillations of the crust is roughly
$10^{48}\alpha^2{\rm~ergs}$ \citep{mcdermott88}. Hence, if the r-mode
could couple efficiently to the crust, it would have adequate energy
to drive crustal pulsations of comparable amplitude and potentially
break the crust.  We now estimate the magnitude of this coupling and
show that, under most circumstances, it should be quite small.

Let us look at a toy problem of a simple harmonic oscillator (i.e.,
the crust) with a mode frequency $\omega_0$ and weak damping
$\gamma\ll\omega_0$ driven by a harmonic force per unit mass $f_0
e^{i\omega t}$.  Away from resonances with crustal pulsation modes,
the response of the crust is then just $\xi_\perp\approx
f_0/(\omega_0^2-\omega^2)$. Ignoring Coriolis force, the fundamental
$l=2$ torsional mode has a frequency $\sim 40\rm{~Hz}$, while the
overtones have frequencies $\sim500n\rm{~Hz}$, where $n\ge1$ is the
radial order of the overtone \citep{mcdermott88}.  Damping times of
these oscillations are on the order of days to decades
\citep{mcdermott88}, so the resonances are narrow.  The spacing
between the crustal modes, $\sim500$~Hz, is comparable to the r-mode
frequency, so on average, $\omega_0^2-\omega^2\sim\omega^2$.  Mass
per unit area in the crust is $M_{\rm{crust}}/4\pi R^2=\rho h$, where
$h\sim1$~km is the scale height at the base of the crust. Therefore,
$f_0=\sigma_{r\perp}/\rho h$ and, in the likely situation that the
r-mode frequency is not in
resonance with a normal mode frequency of the crust, we find 
$\xi_\perp/R\sim\alpha(\delta/h) \approx10^{-4}\alpha/T_8$. 

This simple model (and a preliminary calculation) shows that the
induced toroidal displacements in the crust are much smaller than the
r-mode displacements. In other words, even though the r-mode has
plenty of energy to drive crustal pulsations, it cannot effectively
shake the crust.  Hence, neglecting the coupling should not invalidate
our estimate of the damping time $\tau_{\rm rub}$.

\begin{figure*}[t]
\begin{minipage}[t]{3.5in}
\begin{center}
\epsfig{file=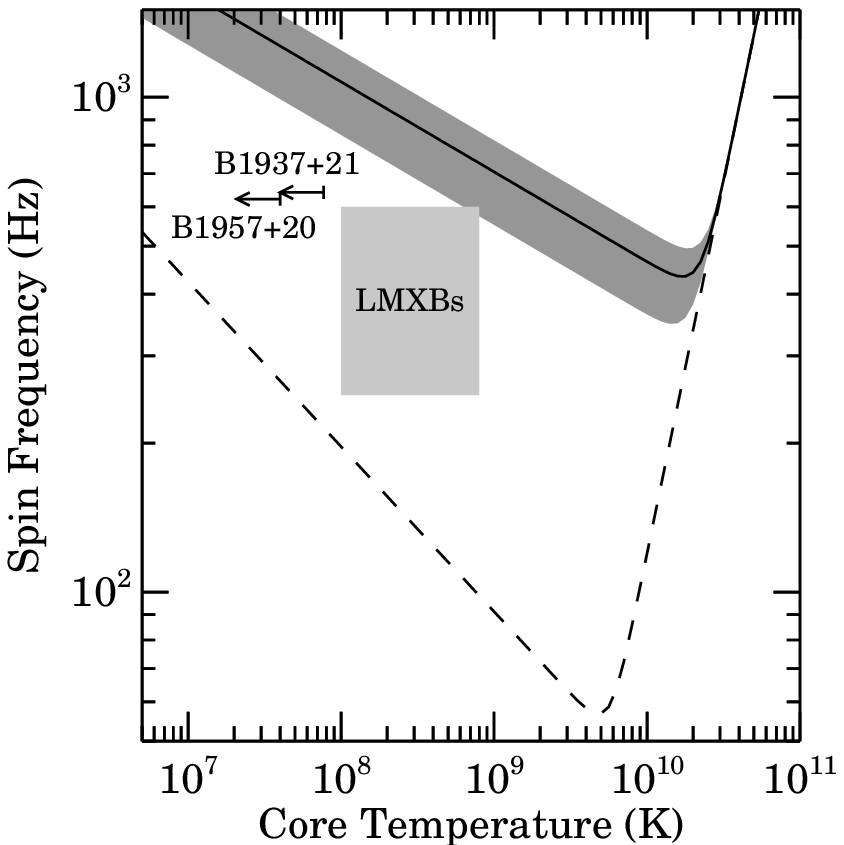}
\end{center}
\vskip -0.15in
\figcaption{
\label{fig:instability} Critical spin frequencies for the
r-mode instability.  Solid line shows the critical frequency set by
the viscous boundary layer and internal dissipation, for a star with
normal nucleons in the core.  The shading around the solid line
displays the effect of core superfluidity.  The dashed line shows the
critical frequency as calculated previously, neglecting the boundary
layer friction.  The LMXBs reside in the rectangular shaded region.
The arrows show the frequencies and the upper limits on core
temperatures (obtained, as in \citealt{andersson99:accreting_rmode}
from \citealt{reisenegger97} and \citealt{gudmundsson82}) of the two
fastest known MSPs.}
\end{minipage}
\hfill
\begin{minipage}[t]{3.5in}
\begin{center}
\epsfig{file=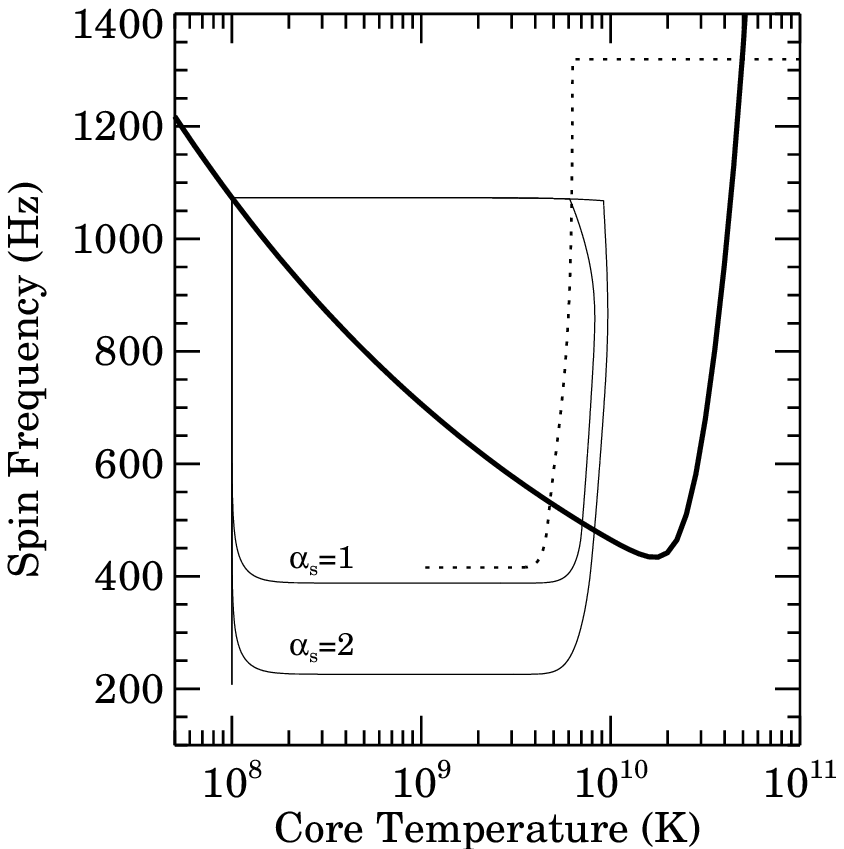}
\end{center}
\vskip -0.15in
\figcaption{
\label{fig:evolution} Evolutionary scenarios for rapidly
rotating neutron stars with crusts.  The thick solid line shows the
critical spin frequency for the r-mode instability, including boundary
layer friction.  The two thin solid loops show the evolution of a
neutron star in an LMXB, accreting at $10^{-8} M_\odot$~yr$^{-1}$, for
two different values of the r-mode saturation amplitude $\alpha_s$, as
marked on the plot.  The dotted line shows the spin-down and cooling of
a newborn rapidly rotating neutron star.}
\end{minipage}
\end{figure*}

\section{The R-mode Instability} 
\label{sec:instability-strip}

We now find the region in $(\nu_s,T)$ space where a NS is unstable to
the r-mode instability by balancing the viscous damping with the
excitation due to gravitational radiation. When the crust is present,
the r-mode is unstable for $\tau_{\rm rub}<|\tau_{\rm gw}|$, or
\begin{equation}\label{eq:nu-min}
\nu_s\gtrsim1070{\rm~Hz~}\frac{M_{1.4}^{4/11}}{R_6^{14/11}}
	\frac{f^{1/11}}{T_8^{2/11}}
	\left(\frac{\rho}{\rhob}\right)^{2/11}.
\end{equation}
However, other viscous mechanisms in the core also contribute to the
damping. In Figure~\ref{fig:instability} we show the critical spin
frequency where $1/\tau_{\rm gw}-1/\tau_{\rm rub}-1/\tau_{\rm
visc}=0$, where we use previously published calculations of shear
\citep{lindblom98:_gravit_radiat_instab} and bulk
\citep{Lindblom99:2ndorder} viscous r-mode damping for the interior
viscous damping time $\tau_{\rm visc}$.  The solid line corresponds to
the case where all nucleons are normal ($f=1$), while the shading
around it represents the range in frequencies when either neutrons or
all nucleons are superfluid.  For comparison, the dashed line shows
the instability curve that neglects the effect of the viscous boundary
layer.  Our curves presume that the crust is present for all
temperatures less than $\sim2\times 10^{10}$~K. For temperatures
higher than that, the bulk viscosity is the dominant damping mechanism.

The shaded box in Figure~\ref{fig:instability} indicates the region of
the $(\nu_s, T)$ space where the observed LMXBs
($250{\rm~Hz}\lesssim\nu_s\lesssim600{\rm~Hz}$,
$T\sim(1-8)\times10^8$~K; \citealt{brown98}) reside. The arrows on the
left-hand side of the plot show frequencies and the upper limits on
the core temperatures for the two fastest MSPs. Clearly, the strength
of the VBL dissipation means that r-modes are not excited in the
accreting and colder ($T\lesssim10^9$~K) NSs.  This result removes the
discrepancy between the existence of $1.6$~ms ($625$~Hz) millisecond
pulsars and the r-mode instability \citep{andersson99:accreting_rmode}
and alleviates the disagreement between the observed and predicted
quiescent luminosities of NS transients \citep{bu99:rmodes}.

 We now re-examine the thermogravitational runaway scenario for LMXBs
\citep{levin99} .  The two closed loops in Figure \ref{fig:evolution}
show the evolution of the spin and temperature of a NS accreting at
$10^{-8} M_\odot$~yr$^{-1}$ (i.e, same as in \citealt{levin99}), but
including the VBL damping.  The evolution is qualitatively the
same. First, the star spins up until it reaches the instability
line. The r-mode amplitude then grows until saturation (at amplitude
$\alpha_s$), heating the star, and then spins it down due to
gravitational wave emission. The r-mode eventually stabilizes and the
star cools down. For the chosen $\alpha_s$, the thermal runaway spins
the star down to $200-400$~Hz.  The final frequency is not sensitive
to the initial temperature of the star, but only to the assumed
$\alpha_s$. The duration of the active r-mode phase is $< \ {\rm yr}$, while
the duration of the cool-down at constant spin is $\sim10^5$~yrs. The
spinup from $\sim400$ to $\sim1100$~Hz takes roughly
$5\times10^6$~yrs. 

Therefore, in agreement with \citet{levin99}, we do
not expect that any of the currently observed LMXBs are in the active
r-mode phase. Moreover, this evolution cannot explain the clustering
of LMXB spins around $300$~Hz \citep{klis99:_millis}, since the
duration of the cool-down at constant spin phase (horizontal leg in
Figure~\ref{fig:evolution}) is much shorter than that of the spin-up
phase (the vertical leg). Other mechanisms, such as mass quadrupole
radiation from the crust \citep{bildsten98:gravity-wave} can likely
explain such clustering.

What about newborn NSs? NSs born in supernovae are thought to rotate
near breakup, and have initial temperatures $T_i\approx10^{11}$~K.
They quickly cool, roughly according to $T_9=(t/\tau_c)^{-1/6}$ for
$T\ll T_i$, where $\tau_c\approx1$~yr is the Urca cooling time at
$10^9$~K \citep{shapiro83}.  During the initial cooling stage, the
r-mode instability line is as calculated previously (dashed line in
Figure~\ref{fig:instability}), and the spin-down evolution proceeds
according to the scenario described by \citet{owen98:_gravit}.
However, a solid crust forms when the NS has cooled to $T_{\rm
m}\sim10^{10}$~K (the exact temperature depends on the composition),
after which we expect the evolution to be significantly altered.  

The dotted line in Figure~\ref{fig:evolution} shows the evolution of
$\nu_s$ and $T$ of a newborn NS, computed as in
\citet{owen98:_gravit}, with initial spin $\nu_{s,i}=1320$~Hz, but
including the VBL and the heating from it.  The star cools from
$T_i=10^{11}$~K and enters the instability region.  The r-mode then
grows, saturates at $\alpha_s=1$, and spins the star down.  We find
that the final spin frequency of the star is $\nu_f=415$~Hz including
the VBL, rather than $120$~Hz \citep{owen98:_gravit}.  The spindown is
also of shorter duration, lasting only $10^4$~s, rather than the
$\sim1$~yr of the original calculations of \citet{owen98:_gravit}.
For shorter Urca cooling time $\tau_c$, the final frequency is
somewhat higher.  The effect of the heating due to the VBL on the
cooling time is negligible.  In our simulation, the r-mode amplitude
grows only by a factor of 3 from its initial value by the time the
star cools to $T_{\rm m}$, so we do not expect the presence of the
r-mode to affect crust formation. In general, the rapid spindown phase
begins when the Urca cooling time at the current temperature,
$T/(dT/dt)\propto T^{-6}$, exceeds the gravitational wave spindown
time, $\Omega/(d\Omega/dt)\propto\alpha_s^{-2}$. Therefore, unless
$T_{\rm m}$ is much less than $10^{10}$~K, the crust will form while
the r-mode amplitude is still rather small.

The strain amplitude is a very shallow function of $\nu_s$ during the
spin-down stage, $h_c\propto\nu_s^{1/2}$ \citep{owen98:_gravit}. 
However, most signal-to-noise ($S/N$) in a
gravitational wave detector is accumulated at low $\nu_s$, where the
source spends most of its time. Hence, the total $S/N$ depends
sensitively on the low-frequency cutoff of the spin-down evolution.
Using the results of \citet{owen98:_gravit}, we find that raising
the cutoff frequency to $415$~Hz from $120$~Hz reduces the $S/N$ from
$\sim8$ to $\sim2.5$ for ``enhanced'' LIGO and a NS at a distance of
$20$~Mpc, where the event rate is expected to be a few per year. 

\section{Conclusions and Future Work}
\label{sec:conclusions}

 Our initial work on the viscous boundary layer between the
oscillating fluid in the core and the co-rotating crust shows that the
dissipation there is very large, making it the predominant r-mode
damping mechanism when the crust is present.  The smallest spin
frequency which allows the r-mode to operate is about 500 Hz, nearly a
factor of five higher than previous estimates. This substantially
reduces the parameter space for the instability to operate, especially
for older, colder NSs, such as those accreting in binaries and
millisecond pulsars.  It resolves the discrepancy between the
theoretical understanding of the r-mode instability and the
observations of millisecond pulsars and LMXBs.

In our estimates we presumed that the boundary layer is laminar.  Of
course, turbulence in the VBL would increase the viscosity there and
move the instability to higher rotation frequencies.  We also showed
that the r-modes induce some transverse motions in the crust, though
these motions appear to be negligibly small compared to the r-mode
amplitude.  The Coriolis force is likely to change the details of the
eigenfunctions in the boundary layer, but it is unlikely to change its
gross structure.  Finally, based on the large conductivity contrast
between the core and the crust, we have presumed that the VBL heating
evenly spreads throughout the isothermal core.  This may be an
oversimplification, and a more detailed time-dependent calculation of
the heat transport will yield the radial temperature profile.  This
is especially important for the accreting systems, where the crust
melting temperatures are lower because of the smaller mean nuclear
charges \citep{haensel90b}.

Can a weak magnetic field substantially change our arguments?  The
Ohmic diffusion time across the VBL is many years, much longer than
the oscillation period. Hence, any magnetic field protruding into the
fluid from the crust will be pulled and sheared in a nearly ideal MHD
response. The field produced by the shear is $\delta B/B\sim\alpha
R/\delta$, and hence the magnetic field does not compete with viscous
shear when $B\ll(\rho\nu\Omega)^{1/2}\approx
10^{11}{\rm~G~}(\nu_s/1{\rm~kHz})^{1/2}/T_8$.  The surface dipole
field strengths in LMXBs and MSPs are lower than this value. If the
same can be said about the $B$ field at the base of the crust, then
the viscous boundary layer we have discussed should apply. Of course,
for younger and potentially much more magnetic neutron stars, the
story will be modified, as the restoring force from the stretched
field lines needs to be self-consistently included.

\acknowledgements 
We thank Andrew Cumming, Yuri Levin, and Lee Lindblom for helpful
conversations.  L.B. is a Cottrell Scholar of the Research Corporation
and G.U. is a Fannie and John Hertz Foundation Fellow. This work was
supported by the National Science Foundation through Grants
AST97-31632 and PHY 94-07194.

\end{document}